\documentclass[12pt]{article}
\usepackage{amssymb,amsmath,epsfig}

\begin{document}

\title{\bf Anisotropic Fluid and Bianchi Type III Model in $f(R)$ Gravity}

\author{M. Sharif \thanks{msharif@math.pu.edu.pk} and H. Rizwana
Kausar
\thanks{rizwa\_math@yahoo.com}\\
Department of Mathematics, University of the Punjab,\\
Quaid-e-Azam Campus, Lahore-54590, Pakistan.}

\date{}
\maketitle

\begin{abstract}
This paper is devoted to study the Bianchi type III model in the
presence of anisotropic fluid in $f(R)$ gravity. Exponential and
power-law volumetric expansions are used to obtain exact solutions
of the field equations. We discuss the physical behavior of the
solutions and anisotropy behavior of the fluid, the expansion
parameter and the model in future evolution of the universe.
\end{abstract}


\section{Introduction}

Hubble diagram of type Ia Supernovae (SNela) measured by the
Supernovae Cosmology Project \cite{super} and High-$z$ Team, up to
the redshift $z\sim1$ \cite{high-z}, has been the first piece of
evidence that current universe is undergoing a phase of accelerated
expansion. The other Balloon-born experiments such as BOOMERanG
\cite{boomer} and MAXIMA \cite{maxima} have detected the anisotropy
spectrum of the Cosmic Microwave Background (CMB) radiations
representing that universe is spatially flat. These data indicate
that present universe is dominated by an un-clustered fluid with
large negative pressure called dark energy (DE) which causes
expansion. The above picture has been further strengthened by the
current measurements of CMB spectrum obtained by Wilkinson Microwave
Anisotropy Probe (WMAP) experiment \cite{cmb},\cite{wmap} and by
extension of the Hubble SNela diagram to redshifts higher than one
\cite{sneIa}. There are various attempts to construct the acceptable
dark energy models in different directions. For example, traditional
cosmological constant, quintessence or phantom models, dark fluid
with complicated equation of state, String or M-theory, higher
dimensions, brane-world models, etc. In addition to these attempts,
the $f(R)$ theory of gravity has also been helpful in describing the
evolution of the universe.

The recent enthusiasm in $f(R)$ theory is caused by its success as
the gravitational alternative for DE \cite{noji1}. Although, the
cosmological constant, $\Lambda$, is the simplest explanation of the
DE and a best fit to some of the astrophysical data \cite{cmb}.
However, the $\Lambda$-Cold Dark Matter ($\Lambda$CDM) fails in
explaining why the value of $\Lambda$ is so tiny (120 orders of
smaller magnitude) as compared to the vacuum energy predicted by
particle physics (the coincidence problem). In order to solve this
problem, people replaced the cosmological constant with a scalar
field referred to quintessence or phantom models \cite{copland}. In
order to explain DE of the current universe with the help of the
above models, one can use the effective equation of state parameter
(EoS) $\omega$. When $w=-1$, the universe passes through
$\Lambda$CDM epoch. If $w<-1$ then we live in the phantom-dominated
universe and for $w>-1$, the quintessence dark era occurs. It is
mentioned here that energy conditions are violated in all these
cases. Also, the phantom phase ends at finite-time singularity in
the future \cite{Mc} while quintessence universe may end up at more
general singularity. This is also true for the $f(R)$ gravity which
leads to DE with the corresponding effective $\omega$ (quintessence
or phantom) \cite{noji1}.

The $f(R)$ theory of gravity quite naturally describes the
transition from deceleration to acceleration in such an evolving
universe \cite{noji-2}. This theory is very useful in high energy
physics, for example, to solve hierarchy or gravity-GUTs unification
problems \cite{cogn}. A large body of papers is available in the
literature \cite{10}-\cite{12} addressing the well-known issues of
gravitational stability \cite{13}, Newtonian limit \cite{14},
singularity problems \cite{15}, solar system test \cite{16}, etc, in
the context of $f(R)$ gravity. The exact solution of the field
equations in metric $f(R)$ theory has been obtained in all the three
symmetric versions. Spherically symmetric solutions have been
investigated for both vacuum and non-vacuum cases by different
authors \cite{sph-sm}. Azadi et al. \cite{cy-s1} and Momeni
\cite{cy-s2} have studied vacuum cylindrically symmetric solutions.
Sharif and Shamir \cite{p-s} have explored plane symmetric solutions
both for vacuum and non-vacuum cases.

It is found that some large-angle anomalies appear in CMB radiations
which violate the statistical isotropy of the universe
\cite{anomalies}. Plane Bianchi models (which are homogeneous but
not necessarily isotropic) seem to be the most promising explanation
of these anomalies. Jaffe et al. \cite{jaffe} investigated that
removing a Bianchi component from the WMAP data can account for
several large-angle anomalies leaving the universe to be isotropic.
Thus the universe may have achieved a slight anisotropic geometry in
cosmological models regardless of the inflation. Further, these
models can be classified according to whether anisotopy occurs at an
early stage or at later times of the universe. The models for the
early stage can be modified in a way to end inflation with a slight
anisotropic geometry \cite{campan}. For the latter class, the
isotropy of the universe, achieved during inflation, can be
distorted by modifying DE \cite{koiv}.

The universe acceleration provides information about the major part
of the universe which has large negative pressure but without
telling anything about the number of cosmic fluids in the universe.
This may be explained by considering the accelerating expansion with
a single fluid and an equation of state acting like DE. The main
benefit of this approach is that a suitable equation of state can be
obtained and observational data can be fitted. In General Relativity
(GR), people have worked by choosing anisotropic fluid with
anisotropic EoS. Akarsu and Kilinc \cite{akar} have studied the
Bianchi type III model in the presence of single imperfect fluid
with dynamical anisotropic EoS parameter. They have concluded that
anisotropy of the DE do not always promote anisotropy of the
expansion. Sharif and Zubair \cite{s-z} have investigated the
Bianchi type $VI_0$ cosmological models in the presence of
electromagnetic field and anisotropic dark energy. They have
examined the effects of electromagnetic field on the dynamics of the
universe. In a recent work \cite{s-r}, we have obtained solutions of
the Bianchi type $VI_0$ universe in $f(R)$ gravity.

Here, our objective is to find exact solutions of the Bianchi type
III model to discuss future evolution of the universe in the
metric $f(R)$ gravity. We take anisotropic fluid to represent DE.
The paper is organized as follows. In section \textbf{2}, we
present some features of the Bianchi type III model and define
dynamical quantities describing the evolution of the universe.
Section \textbf{3} provides the field equations and anisotropic
parameter of the expansion. We obtain possible solutions of the
field equation and discuss the physical behavior of these
solutions in section \textbf{4}. In the last section \textbf{5},
we summarize the results obtained.

\section{The Model}

The homogeneous and anisotropic Bianchi type III spacetime is
described by the line element
\begin{equation}\label{1}
ds^{2}=dt^2-A^2(t)dx^2-e^{-2\alpha x}B^2(t)dy^2-C^2(t)dz^2,
\end{equation}
where the scale factors $A,~B$ and $C$ are only functions of
cosmic time $t$ and $\alpha\neq0$ is a constant.

The energy-momentum tensor for anisotropic fluid is given as
\begin{eqnarray}\label{2}
T_{\mu}^{\nu}&=&diag[\rho,-p_{x},-p_{y},-p_{z}]=diag[1,-{\omega}_{x}
,-{\omega}_{y},-{\omega}_{z}]\rho,
\end{eqnarray}
where $\rho$ is the density of the fluid while $p_x$, $p_y$ and
$p_z$ are pressures and $\omega_x$, $\omega_y$ and $\omega_z$ are
directional EoS parameters on the $x$,$y$ and $z$ axes
respectively. The DE schematically characterized by this
parameter. The deviation from isotropy may be obtained by setting
\begin{equation*}
{\omega}_x={\omega}, \quad {\omega}_y={\omega}+{\delta} \quad
\textmd{and} \quad {\omega}_{z}={\omega}+{\gamma},
\end{equation*}
where $\omega$ is the deviation-free EoS parameter and $\delta$ and
$\gamma$ are the deviations from $\omega$ on $y$ and $z$ axes
respectively. In this case, the energy-momentum tensor becomes
\begin{equation}\label{12}
T_{\mu}^{\nu}=diag[1,-{\omega},-({\omega}+{\delta}),-({\omega}+{\gamma})]\rho.
\end{equation}
The skewness parameters $\delta$ and $\gamma$ may be constant or
functions of cosmic time. The average scale factor and volume of the
universe for this model will be
\begin{equation}\label{24}
a=(ABC)^{1/3}, \quad V=a^3=ABC.
\end{equation}
The mean Hubble parameter $H$, deceleration parameter $q$, expansion
scalar $\Theta$ and the directional Hubble parameters in the $x$,
$y$ and $z$ directions turn out to be
\begin{eqnarray}\label{26}
&&H=\frac{\ln\dot{V}}{3}=\frac{1}{3}(\frac{\dot{A}}{A}+\frac{\dot{B}}{B}
+\frac{\dot{C}}{C}),\quad
q=-\frac{a\ddot{a}}{\dot{a}^2}, \\
&&\Theta=u_{;a}^a=\frac{\dot{A}}{A}+\frac{\dot{B}}{B}+\frac{\dot{C}}{C},\quad
H_x=\frac{\dot{A}}{A},\quad H_y=\frac{\dot{B}}{B},\quad
H_z=\frac{\dot{C}}{C}.
\end{eqnarray}
Here the deceleration parameter $q$ measures the rate of expansion
of the universe. The sign of $q$ indicates the state of expanding
universe. If $q<0$ or $q>0$ respectively, then it represents
inflation or deflation of the universe while $q=0$ shows expansion
with constant velocity.

To examine whether expansion of the universe is anisotropic or not,
we define anisotropic expansion parameter as
\begin{equation}\label{aniso}
\Delta=\frac{1}{3}\sum_{i=1}^3(\frac{H_i-H}{H})^2,
\end{equation}
where $H_i$ $(i=1,2,3)$ are the directional Hubble parameters in
the directions of $x,y$ and $z$ axes respectively. If $\Delta=0$,
then the universe expands isotropically. Further, any anisotropic
model of the universe with diagonal energy-momentum tensor
approaches to isotropy if $\Delta\rightarrow0$,
$V\rightarrow{+\infty}$ \textmd{and} $\rho>0$ as
$t\rightarrow{+\infty}$ (\cite{akar},\cite{collins}).

\section{The Field Equations in $f(R)$ Gravity}

The Lagrangian in $f(R)$ gravity is given as \cite{1*}
\begin{equation}\setcounter{equation}{1}\label{3.1}
L=\frac{f(R)}{2\kappa}+L_{M},
\end{equation}
where $f(R)$ is a function of the Ricci scalar and $L_{M}$ describes
all kinds of matter including non-relativistic (cold) dark matter.
Assuming variation with respect to the metric tensor, $g_{\mu\nu}$,
we have the following fourth order partial differential equations
\begin{equation}\label{3.2}
F(R) R_{\mu\nu} - \frac{1}{2}f(R)g_{\mu\nu}-\nabla_{\mu}
\nabla_{\nu}F(R)+ g_{\mu\nu} \Box F(R)= \kappa T_{\mu\nu},
\end{equation}
where $F(R)\equiv df(R)/dR,~\Box \equiv \nabla^{\mu}\nabla_{\mu}$
with $\nabla_{\mu}$ representing the covariant derivative and
$\kappa(=\frac{8\pi G}{c^4}=1)$ is the coupling constant in
gravitational units. The trace of Eq.(\ref{3.2}) yields
\begin{equation}\label{3.3}
F(R) R - 2f(R)+ 3\Box F(R)= T.
\end{equation}
This equation is helpful to solve the field equations and also to
express $f(R)$ in terms of its derivative, i.e.,
\begin{equation}\label{3.4}
f(R)=\frac{-T+ F(R) R+ 3\Box F(R)}{2}.
\end{equation}
The scalar curvature for Bianchi type III model is given by
\begin{equation}\label{3.R}
R=-2\left[\frac{\ddot{A}}{A}+\frac{\ddot{B}}{B}+\frac{\ddot{C}}{C}-\frac{\alpha^2}{A^2}
+\frac{\dot{A}\dot{B}}{AB}+\frac{\dot{A}
\dot{C}}{AC}+\frac{\dot{B} \dot{C}}{BC}\right].
\end{equation}

The corresponding field equations become
\begin{eqnarray}\label{3.5}
(\frac{\ddot{A}}{A}+\frac{\ddot{B}}{B}+\frac{\ddot{C}}{C})F
+\frac{1}{2}f(R)-(\frac{\dot{A}}{A}+\frac{\dot{B}}{B}+\frac{\dot{C}}{C})\dot{F}
=-\rho,&&\\\label{3.6}
(\frac{\ddot{A}}{A}-\frac{\alpha^2}{A^2}+\frac{\dot{A}\dot{B}}{AB}+\frac{\dot{A}
\dot{C}}{AC})F
+\frac{1}{2}f(R)-\ddot{F}-(\frac{\dot{B}}{B}+\frac{\dot{C}}{C})\dot{F}
=-\omega\rho,&&\\\label{3.7}
(\frac{\ddot{B}}{B}-\frac{\alpha^2}{A^2}+\frac{\dot{A}\dot{B}}{AB}+\frac{\dot{B}
\dot{C}}{BC})F
+\frac{1}{2}f(R)-\ddot{F}-(\frac{\dot{A}}{A}+\frac{\dot{C}}{C})\dot{F}
=-(\omega+\delta)\rho,&&\\\label{3.8}
(\frac{\ddot{C}}{C}+\frac{\dot{A}\dot{C}}{AC}+\frac{\dot{B}\dot{C}}{BC})F
+\frac{1}{2}f(R)-\ddot{F}-(\frac{\dot{A}}{A}+\frac{\dot{B}}{B})\dot{F}
=-(\omega+\gamma)\rho,&&\\\label{3.9}
\alpha(\frac{\dot{A}}{A}-\frac{\dot{B}}{B})F=0.&&
\end{eqnarray}
The solution of Eq.(\ref{3.9}) yields
\begin{equation}\label{3.10}
B=c_1A,
\end{equation}
where $c_1$ is a constant of integration. Using Eq.(\ref{3.10}) and
then subtracting Eq.(\ref{3.7}) from Eq.(\ref{3.8}), we obtain
$\delta=0$. This indicates that directional EoS parameters
$\omega_x,~\omega_y$ along $x$ and $y$ axes become equal and hence
pressures. Consequently, the field equations turn out to be
\begin{eqnarray}\label{3.12}
&&(\frac{2\ddot{A}}{A}+\frac{\ddot{C}}{C})F
+\frac{1}{2}f(R)-(\frac{2\dot{A}}{A}+\frac{\dot{C}}{C})\dot{F}
=-\rho,\\\label{3.13}
&&(\frac{\ddot{A}}{A}-\frac{\alpha^2}{A^2}+\frac{\dot{A}^2}{A^2}+\frac{\dot{A}\dot{C}}{AC})F
+\frac{1}{2}f(R)-\ddot{F}-(\frac{\dot{A}}{A}+\frac{\dot{C}}{C})\dot{F}=-\omega\rho,
\\\label{3.14}
&&(\frac{\ddot{C}}{C}+\frac{2\dot{A}\dot{C}}{AC})F
+\frac{1}{2}f(R)-\ddot{F}-(\frac{2\dot{A}}{A})\dot{F}=-(\omega+\gamma)\rho.
\end{eqnarray}
Subtracting Eq.(\ref{3.13}) from Eq.(\ref{3.14}) and integrating the
resulting equation, we have
\begin{equation}\label{3.15}
\frac{\dot{A}}{A}-\frac{\dot{C}}{C}=\frac{c}{VF}+
\frac{1}{VF}\int{(\frac{\alpha^2F}{A^2}+\gamma\rho)Vdt},
\end{equation}
where $c$ is a positive constant of integration. In terms of
directional Hubble parameters, the above equation can be written as
\begin{equation}\label{3.16}
H_x-H_z=\frac{c}{VF}+
\frac{1}{VF}\int{(\frac{\alpha^2F}{A^2}+\gamma\rho)Vdt}.
\end{equation}
Using Eqs.(\ref{3.16}) and (\ref{aniso}), it follows that
\begin{equation}\label{3.17}
\Delta=\frac{2}{9H^2}[c+\int{(\frac{\alpha^2F}{A^2}+\gamma\rho)}Vdt]^2V^{-2}F^{-2}.
\end{equation}
If we take $\gamma=0$ and $F(R)=1$, the anisotropy parameter of
expansion reduces to GR for an isotropic fluid
\begin{equation}\label{3.18}
\Delta=\frac{2}{9H^2}[c+\int{\frac{\alpha^2FV}{A^2}}dt]^2V^{-2}F^{-2}.
\end{equation}
The integral in Eq.(\ref{3.17}) vanishes for the following value of
$\gamma$ \cite{akar,s-z}, i.e.,
\begin{equation}\label{3.19}
\gamma=-\frac{\alpha^2F}{\rho A^2}.
\end{equation}
Consequently, the energy-momentum tensor, anisotropy parameter and
Eq.(\ref{3.16}) will take the form
\begin{eqnarray}\label{3.20}
T_{\mu}^{\nu}&=&diag[1,-\omega,-{\omega},-\omega+\frac{\alpha^2F}{\rho
A^2}]\rho,\\\label{3.21}
\Delta&=&\frac{2}{9}\frac{c^2}{H^2}V^{-2}F^{-2},\\\label{3.22}
H_x-H_z&=&\frac{c}{VF}.
\end{eqnarray}
It is interesting to mention here that these results will reduce to
GR \cite{akar} for $F(R)=1$.

\section{Solutions of the Field Equations}

In this section we would obtain exact solutions of the Bianchi type
III model in the presence of anisotropic fluid. Using
Eq.(\ref{3.19}) in Eqs.(\ref{3.12})-(\ref{3.14}) along with
Eq.(\ref{3.3}), we have
\begin{eqnarray}\setcounter{equation}{1}\label{4.1}
(\frac{2\ddot{A}}{A}+\frac{\ddot{C}}{C})F
+\frac{1}{2}f(R)-(\frac{2\dot{A}}{A}+\frac{\dot{C}}{C})\dot{F}
=-\rho,&&\\\label{4.2}
(\frac{\ddot{A}}{A}+\frac{\dot{A}^2}{A^2}+\frac{\dot{A}\dot{C}}{AC})F
+\frac{1}{2}f(R)-\ddot{F}-(\frac{\dot{A}}{A}+\frac{\dot{C}}{C})\dot{F}=
-(\omega+\gamma)\rho,&&
\\\label{4.3}
(\frac{\ddot{C}}{C}+\frac{2\dot{A}\dot{C}}{AC})F
+\frac{1}{2}f(R)-\ddot{F}-(\frac{2\dot{A}}{A})\dot{F}=-(\omega+\gamma)\rho,&&\\\nonumber
-(\frac{2\ddot{A}}{A}+\frac{\ddot{C}}{C}-\frac{\alpha^2}{A^2}+\frac{\dot
{A}^2}{A^2}+\frac{2\dot{A}\dot{C}}{AC})F
-2f(R)+3\ddot{F}&&\\\label{trace}
+3(\frac{\dot{2A}}{A}+\frac{\dot{C}}{C})\dot{F}=(1-3\omega-\gamma)\rho.&&
\end{eqnarray}
This is a system of four differential equations in five unknowns. In
order to obtain solution, we make use of two different volumetric
expansion laws (in next subsections) to get a system of five
equations with five unknowns.

\subsection{Bianchi Type III Model for Exponential Expansion}

We consider the following volumetric exponential expansion law
\begin{equation}\label{4.4}
V=c_2e^{3kt},
\end{equation}
where $c_2$ and $k$ are positive constants. Using the value of $V$
in Eq.(\ref{3.22}), we have
\begin{equation}\label{4.5}
H_x-H_z=\frac{c e^{-3kt}}{c_2F}.
\end{equation}
Solving the system of equations (\ref{4.1})-(\ref{trace}) and
Eq.(\ref{4.5}), we obtain the scale factors as follows
\begin{eqnarray}\label{4.6}
A&=&c_3e^{kt+\frac{c}{3c_2}\int{\frac{e^{-3kt}}{F}dt}},\\\label{4.7}
B&=&c_1c_3e^{kt+\frac{c}{3c_2}\int{\frac{e^{-3kt}}{F}dt}},\\\label{4.8}
C&=&c_4e^{kt-\frac{2c}{3c_2}\int{\frac{e^{-3kt}}{F}dt}},
\end{eqnarray}
where $c_3$ and $c_4$ are positive constants of integration. To
solve the integral part in the above equations, we assume a
relation between $F$ and $V$ as $F\alpha V^{m/3}$ \cite{p-s},
which gives
\begin{equation}\label{4.9}
F=c_5V^{m/3}\quad\Rightarrow\quad F=c_5 e^{mkt},
\end{equation}
where $c_5$ is a positive proportionality constant and $m$ is a
non-zero arbitrary constant. Using Eq.(\ref{4.9}) in
Eqs.(\ref{4.6})-(\ref{4.8}), the scale factors become
\begin{eqnarray}\label{4.10}
A&=&c_3e^{kt-\frac{c}{3c_2c_5}\frac{1}{k(3+m)}e^{-(3+m)kt}},\\\label{4.11}
B&=&c_1c_3e^{kt-\frac{c}{3c_2c_5}\frac{1}{k(3+m)}e^{-(3+m)kt}},\\\label{4.12}
C&=&c_4e^{kt+\frac{2c}{3c_2c_5}\frac{1}{k(3+m)}e^{-(3+m)kt}}.
\end{eqnarray}
This is the first solution of Bianchi type III model with
exponential volumetric expansion. When $m>-3$, the scale factors
admit constant values at early times of the universe, after that
start increasing with the increase in cosmic time without showing
any type of initial singularity and finally diverges to $\infty$
for $t\rightarrow\infty$. This shows that at the initial epoch,
the universe starts with zero volume and expands exponentially
approaching to infinite volume. However, for $m<-3$, the scale
factors $A,~B$ increase with time while $C$ approaches to zero as
$t\rightarrow\infty$. Also, the model with $m=-3$ represents the
universe at early stage. Moreover, the expansion scalar for these
scale factors exhibits the constant value, i.e., $\Theta=3k$ which
shows uniform exponential expansion from $t=0$ to $t=\infty$.

The mean, directional and deceleration parameters turn out to be
\begin{eqnarray}\label{4.13}
H_x&=&H_y=k+\frac{c}{3c_2c_5}e^{-(3+m)kt}, \quad
H_z=k-\frac{2c}{3c_2c_5}e^{-(3+m)kt},\nonumber\\
H&=&k, \quad q=-1.
\end{eqnarray}
This shows that the mean Hubble parameter is constant whereas others
are dynamical. As time approaches from zero to infinity (for
$m>-3$), the directional Hubble parameters start reducing towards
the constant value of $H$ and becomes equal as $t\rightarrow\infty$.
For $m<-3$, parameters along $x$ and $y$ axes will increase from the
mean Hubble parameter by a constant factor $\frac{c}{3c_2c_5}$ while
parameter along $z$-axis decreases by twice the same factor. Also,
the deceleration parameter appears with negative sign which shows
accelerating expansion of the universe as we can expect for
exponential volumetric expansion.

The anisotropy parameter of the expansion takes the form
\begin{equation}\label{4.14}
\Delta=\frac{2}{9}(\frac{c}{kc_2c_5})^2e^{-2(m+3)kt}.
\end{equation}
Here one can observe that at $t=0$, the anisotropy parameter
measures a constant value while it vanishes for $m>-3$ at infinite
time of the universe. This indicates that the universe expands
isotropically at later times without taking any effect from
anisotropy of the fluid. However, for $m<-3$, the anisotropy in the
expansion will increase with time.

Inserting the scale factors, Eqs.(\ref{4.10})-(\ref{4.12}), in
Eqs.(\ref{4.1})-(\ref{trace}) and use Eq.(\ref{4.9}), we obtain the
following energy density of anisotropic dark fluid
\begin{eqnarray}\nonumber
\rho&=&\frac{c_5e^{mkt}}{3}[-3m^2k^2+2mk^2+\frac{2}{3}\frac{mkc}{c_2c_5}e^{-(3+m)kt}
-\frac{2c^2}{(c_2c_5)^2}e^{-2(3+m)kt}\\\label{4.15}
&-&\frac{2\alpha^2}{c_3^2}e^{-2kt+\frac{2c}{3c_2c_5}\frac{e^{-(3+m)kt}}{k(3+m)}}].
\end{eqnarray}
We see that the matter density is constant at early stage of the
universe ($t=0$) and shows monotonic behavior in the evolving cosmic
times. This behavior of matter density deviates from GR where it
becomes constant in exponential expansion. Here, for positive values
of $m$, it will positively increase with time and diverges to
$\infty$, showing future time singularity. For $-3<m<0$, it starts
decreasing with cosmic time and eventually approaches to zero as
$t\rightarrow\infty$. It is mentioned here that isotropic conditions
of the model may not be satisfied and hence leading to the
conclusion that Bianchi type III model remains anisotropic in the
later times whereas in GR it becomes isotropic. The anisotropy of
the model is maintained not due to the presence of anisotropic fluid
but due to modification in gravity. It is worth mentioning here that
the results correspond to GR results for $F(R)=1$.

The anisotropic EoS parameter obtained from the field equations is
given by
\begin{eqnarray}\nonumber
\omega=&&[3m(kc_2c_3c_5)^2-2mkc
c_2c_5e^{-(3+m)kt}+9\alpha^2c_2^2c_5^2e^{-2kt+
\frac{2c}{3c_2c_5}\frac{e^{-(3+m)kt}}{k(3+m)}}]/\\\nonumber
&&[3m(kc_2c_3c_5)^2(-3m+2)+2mkc
c_2c_5e^{-(3+m)kt}-6c^2c_3^2e^{-2(3+m)kt}\\\label{4.16}
&&-6\alpha^2c_2^2c_5^2e^{-2kt+\frac{2c}{3c_2c_5}\frac{e^{-(3+m)kt}}{k(3+m)}}].
\end{eqnarray}
When $t\rightarrow\infty$ this gives a constant quantity,
$\frac{1}{2-3m}$, which can further be characterized into phantom
and quintessence regions for $m>1$ or $-3<m<1$ respectively. For
$m=1$, the EoS parameter is $\omega=-1$, corresponds to the vacuum
energy which is mathematically equivalent to cosmological constant
while $m<-3$ does not yield fruitful results. We see that $\rho$
increases rapidly in the phantom region where $\omega<-1$. It is
mentioned here that phantom regime support the observational
evidence of a recent supernova data \cite{A}-\cite{S}. However, it
decreases in the quintessence region representing relatively slow
expansion.

Using Eqs.(\ref{4.9}), (\ref{4.10}) and (\ref{4.15}) in
Eq.(\ref{3.19}), the skewness parameter along $z$-axis becomes
\begin{eqnarray}\nonumber
\gamma=&&[-9\alpha^2c_2^2c_5^2e^{-2kt+
\frac{2c}{3c_2c_5}\frac{e^{-(3+m)kt}}{k(3+m)}}]/\\\nonumber
&&[3m(kc_2c_3c_5)^2(-3m+2)+2mkc
c_2c_5e^{-(3+m)kt}-6c^2c_3^2e^{-2(3+m)kt}\\\label{4.17}
&&-6\alpha^2c_2^2c_5^2e^{-2kt+\frac{2c}{3c_2c_5}\frac{e^{-(3+m)kt}}{k(3+m)}}].
\end{eqnarray}
We have already found $\delta=0$ and we see that $\gamma$ also
approaches to zero as $t\rightarrow\infty$. Thus the anisotropy of
the fluid is completely removed in the future evolution of the
universe. The scalar curvature and $f(R)$ function for exponential
model becomes
\begin{eqnarray}\nonumber
R&=&-10k^2-\frac{4}{9}\frac{c^2}{c_2^2c_5^2}e^{-2(3+m)kt}+
\frac{4}{3}\frac{kc}{c_2c_5}e^{-(3+m)kt}+ \frac{2\alpha^2}{c_3^2}
e^{-2kt+\frac{2ce^{-(3+m)kt}}{3c_2c_5k(3+m)}}\\\nonumber
f(R)&=&\frac{c_5e^{mkt}}{3}[(36m^2k^2+84mk^2+9R)c_2^2c_2^2c_5^2-4mk
\lambda c_2c_5e^{-2(3+m)kt}\\\label{4.19}
&&+6\lambda^2c_3^2e^{-2(3+m)kt}+24\alpha^2c_2^2c_5^2
e^{-kt+\frac{2\lambda}{3c_2c_5}\frac{e^{-(3+m)kt}}{k(3+m)}}].
\end{eqnarray}
This becomes constant in the quintessence region when
$t\rightarrow\infty$ while for $m<-3$, it diverges giving the same
behavior as the energy density. Since the energy density decreases
in the quintessence region, thus the universe starts from inflation
driven by the anisotropic energy density at the early stage where
curvature is very large. The scalar curvature also reduces with the
passage of time. After that time, the matter density or radiations
become small and the curvature becomes constant.

\subsection{Bianchi Type III Model for Power-law Expansion}

Here we solve the field equations by assuming power-law volumetric
expansion given as
\begin{equation}\label{5.1}
V=c_2t^{3n},
\end{equation}
where $c_2$ and $n$ are positive constants. For $n>1$, Bianchi
models exhibit accelerating volumetric expansion. When $n=1$, the
models represent volumetric expansion with constant velocity while
for $n<1$ these show decelerating volumetric expansion. Adopting the
same procedure as in the exponential model, we obtain the scale
factors
\begin{eqnarray}\label{5.4}
A&=&c_3t^n e^{\frac{c}{3c_2c_5(1-3n-mn)}t^{1-3n-mn}},\\\label{5.5}
B&=&c_1c_3t^n
e^{\frac{c}{3c_2c_5(1-3n-mn)}t^{1-3n-mn}},\\\label{5.6} C&=&c_4t^n
e^{\frac{-2c}{3c_2c_5(1-3n-mn)}t^{1-3n-mn}}.
\end{eqnarray}
We discuss the behavior of these scale factors in the future
evolution by taking very large values of time. It is found that
for $n<\frac{1}{3+m}$, the scale factors $A$ and $B$ increase with
time while $C$ approaches to zero provided that $m$ is always be
greater than $-3$ because $n$ is assumed to be a positive number.
This indicates that the universe is expanding along $x$ and $y$
axes with deceleration whereas there is no expansion along
$z$-axis. Similarly, for $n>\frac{1}{3+m}$, the behavior of scale
factors interchanged, however, the range $-3<m\leq-2$ lies in the
accelerating phase and $-2<m$ comprises the decelerating epoch.

The Hubble parameters, scalar expansion and anisotropy parameter of
expansion in this case become
\begin{eqnarray}\label{5.7}
H&=&\frac{n}{t},\quad H_x=\frac{n}{t}+\frac{c
t^{-(3+m)n}}{3c_2c_5},\quad
H_z=\frac{n}{t}-\frac{2c}{3c_2c_5}t^{-(3+m)n},\\\label{5.8}
\Theta&=&\frac{3n}{t},\quad
\Delta=\frac{2}{9}(\frac{c}{nc_2c_5})^2t^{2(1-3n-mn)}.
\end{eqnarray}
We see that the Hubble parameters and scalar expansion are extremely
large at the origin of the universe and start decreasing
monotonically with the passage of time and possibly will take zero
value in the future (provided that $m>-3$). This shows that, at
earlier times of the universe (just after the big bang), the
expansion is much faster but slows down for later time of the
universe. The anisotropy parameter approaches to zero for
$n>\frac{1}{3+m}$ leaving isotropic expansion in the future while
for $n<\frac{1}{3+m}$, its behavior is switched.

With the same manipulations as in the previous subsection, the
unknown functions in the field equations turn out to be
\begin{eqnarray}\nonumber
\rho&=&\frac{c_5t^{mn}}{3}[(-3m^2n^2+3mn+2mn^2+6n)t^{-2}+\frac{2}{3}\frac{mnc}{c_2c_5}
t^{-(3+m)n-1}\\\label{5.9} &&-(\frac{2c^2}{c_2c_5})^2t^{-2(3+m)n}
-\frac{2\alpha^2}{c_3^2}t^{-2n}e^{\frac{-2c}{3c_2c_5}\frac{t^{1-3n-mn}}{(1-3n-mn)}}],
\\\nonumber
\omega&=&[9\alpha^2c_2^2c_5^2t^{-2n}e^{\frac{-2c}{3c_2c_5}
\frac{t^{1-3n-mn}}{(1-3n-mn)}}-2c_2c_5c_3^2c mn
t^{-(3+m)n-1}+3mn^2(c_2c_3c_5)^2t^{-2}]\\\nonumber &&/
[3(c_2c_3c_5)^2(-3mn(mn-1)+2mn^2+6n)t^{-2}+2mnc
c_2c_5c_3^2t^{-(3+m)n-1}\\\label{5.10}
&&-6c^2c_3^2t^{-2(3+m)n}-6\alpha^2c_2^2c_5^2t^{-2n}e^{\frac{-2c}
{3c_2c_5}\frac{t^{1-3n-mn}}{(1-3n-mn)}}],\\\nonumber
\gamma&=&[-9\alpha^2c_2^2c_5^2t^{-2n}e^{\frac{-2c}{3c_2c_5}
\frac{t^{1-3n-mn}}{(1-3n-mn)}}]/[2(c_2c_3c_5)^2(-3mn(mn-1)+2mn^2\\\nonumber
&&+6n)t^{-2}+2mnc
c_2c_5c_3^2t^{-(3+m)n-1}-6c^2c_3^2t^{-2(3+m)n}\\\label{5.11}
&&-6\alpha^2c_2^2c_5^2t^{-2n}e^{\frac{-2c}{3c_2c_5}\frac{t^{1-3n-mn}}{(1-3n-mn)}}].
\end{eqnarray}
In the accelerating models, the energy density will increase with
times while for the decelerating models it will decrease. The dark
energy EoS parameter for the accelerating models gives
$\omega=-1+\frac{6-3m(mn-1)+3mn}{6-3m(mn-1)+2mn}$ for later times.
This indicates that the universe will pass through quintessence
epoch with appropriate values of $m>-3$. The skewness deviation
parameter along $z$-axis vanish in the future evolution. Thus DE
isotropous in the far future. The scalar curvature is
\begin{eqnarray}\label{5.12}\nonumber
R&=&(6n-10n^2)t^{-2}-\frac{32}{9}\frac{c^2}{c_2^2c_5^2}t^{-2(3+m)n}+
\frac{4c n}{3c_2c_5}t^{-(3+m)n}\\ &+&\frac{2\alpha^2}{c_3^2}
t^{-2n}e^{\frac{-2c}{3c_2c_5}\frac{t^{1-3n-mn}}{(1-3n-mn)}},\\\nonumber
f(R)&=&\frac{c_5t^{mn}}{2}[(4mn(mn-1)+\frac{25}{3}mn^2+6n)t^{-2}+R\\\nonumber
&+& \frac{8}{3}\frac{\alpha^2}{c_3^2}
t^{-2n}e^{\frac{-2\lambda}{3c_2c_5}\frac{t^{1-3n-mn}}{(1-3n-mn)}}
+\frac{2\lambda^2}{3c_3}t^{-2(3+m)n}-\frac{2\lambda
mn}{9c_2c_5}t^{-(3+m)n-1}].\\
\end{eqnarray}
The scalar curvature approaches to zero for the passing times
similar to $\rho$.

\section{Summary and Outlook}

The aim of this paper is to discuss expansion of the universe due
to anisotropic DE in $f(R)$ gravity. For this purpose, we have
found exact solutions of Bianchi type III model by assuming
exponential and power-law volumetric expansion to cover all types
of the expansion histories. These models represent an accelerated
expansion of the universe with $V\rightarrow\infty$ as
$t\rightarrow\infty$ and supports the observations of the Type Ia
supernova \cite{super,high-z} and WMAP data \cite{cmb,wmap}. The
whole discussion is made in terms of $F(R)$ which is assumed to
have direct proportion with volume of the universe. The results of
the paper can be summarized as follows:

\begin{itemize}

\item The solution of the field equations for Bianchi type III model
yields that the scale factors $A$ and $B$ are equal and the
deviation from isotropy along $y$-axis is zero (i.e., $\delta=0$).
This shows that pressure of DE along $x$ and $y$ axes is same.

\item The physical behavior of the dynamical quantities depends on the
values of $m$. In case of exponential model, the dynamical
parameters show different behavior for $m\gtrless-3$ while in
power-law model we can only discuss $m>-3$ for later times to keep
$n$ positive. However, here we have further two types of models for
$n<\frac{1}{3+m}$ or $n>\frac{1}{3+m}$.

\item The scalar expansion is constant for the first solution
indicating that the universe expands homogeneously whereas in the
second solution it shows that the expansion rate is faster at the
beginning but slows down with the passage of time. The anisotropy
parameter of expansion at $t=0$ measures a constant value while it
vanishes for $m>-3$ at infinite time of the universe. This shows
that the universe expands isotropically at later times. It is
interesting to mention here that our solution can be used to
understand cosmological phase transitions. It may help to
understand the isotropization mechanism which could be responsible
for the passage from a possible prior anisotropic phase to the
present isotropic era we live in.

\item Behavior of EoS parameter and $\rho$ exhibit that the universe will pass through
quintessence region and phantom regions for $m>1$ or $-3<m<1$
respectively. It is worth mentioning here that solution in the
phantom regime ($\omega<-1$) supports the observational evidence of
a recent supernova data \cite{A}-\cite{S}. Notice that the energy
density increases/decreases monotonically with the passage of time
depending upon the value of $m$ in $f(R)$ gravity. However, it
becomes constant in GR at later times. The skewness parameter along
$z$-axis will vanish thus acquiring isotropy in the fluid at
infinite times.

\item The Bianchi type III model remains anisotropic even in the later
times which is different from GR \cite{akar} where these become
isotropic.

\item The scalar curvature becomes constant for $m>-3$ and diverges for $m<-3$
in exponential models while it becomes zero in the power-law models
for later times.
\end{itemize}

\vspace{0.5cm}

{\bf Acknowledgment}

\vspace{0.25cm}

We would like to thank the Higher Education Commission, Islamabad,
Pakistan for its financial support through the {\it Indigenous
Ph.D. 5000 Fellowship Program Batch-III}.

\end{document}